\documentclass[letterpaper]{sfchem}
\usepackage{graphicx}

\newcommand\araa{{ARA\&A}}%
\newcommand\aj{{AJ}}%
\newcommand\apj{{ApJ}}%
\newcommand\apjl{{ApJ}}%
\newcommand\apjs{{ApJS}}%
\newcommand\apss{{Ap\&SS}}%
\newcommand\aap{{A\&A}}%
\newcommand\qjras{{QJRAS}}%
\newcommand\mnras{{MNRAS}}%
\newcommand\nat{{Nature}}%
\def\HCOp{HCO$^+$}             % HCO+
\def\CeiO{C$^{18}$O}           % C18O
\def\NtwoHp{N$_2$H$^+$}        % N2H+ 
\def\NHthree{NH$_{3}$}         % NH3

\begin{document}

\title{The Chemistry of Dark Clouds: New Astrochemical Tools for Star Formation
Studies}

\author{Edwin A. Bergin\inst{1}
}
  \institute{Harvard Smithsonian Center for Astrophysics, 60 Garden St., Cambridge MA
02138}
%Short author list here: Surnames only please (no initials)
\authorrunning{Bergin}
%Short title here:
\titlerunning{Chemistry and Star Formation}

\maketitle 

\begin{abstract}
The past decade has led to significant improvements in our understanding of the physical
structure of the molecular cores of cold dark
clouds. Observational efforts, in combination with improved knowledge of cloud structure,
now provide clear evidence that the
chemistry of dark clouds is dominated by the depletion of gaseous species onto grain
surfaces. We outline the basis of these
observational efforts and show how the abundance determinations have moved beyond single
point analyses to the derivation of abundance profiles.
We discuss the basic physics of the interaction between
molecules and grain surfaces and show that when 
physics is coupled into a chemical model there is excellent agreement, for a limited set
of species, between theory and observations.
We discuss our improved understanding of cloud chemistry 
can be used as a new tool for studies
of the formation of stars and planetary systems. 
\end{abstract}

\section{Introduction}

The formation of dense molecular condensations, and eventually stars, 
involves large changes in the physical properties of the atomic
and molecular gas.   These changes also have very specific
consequences on the chemical interactions of the gas and dust inside the forming
cores/stars.   In particular as the density increases molecules in the
gas phase collide with dust grains with greater frequency and, if the
molecules stick with any reasonable efficiency, they will deplete from the  
gas phase.  
Thus, the density gradient that results from core condensation
is accompanied by chemical gradients, with the inner parts of the
core representing high density chemistry (with freeze-out onto grains), and
the outer parts representing the original low density molecular composition.
Indeed, these molecular depletions have been suggested as the 
primary chemical indicator of the earliest stages of the star formation process 
(Mundy \& McMullin 1997; Bergin \& Langer 1997).   Such chemical structure
should be particularly evident in dark clouds because the low temperatures
($\le 10$ K) preclude thermal evaporation of the molecules frozen in the mantle. 

The process of star formation is by nature a dynamic one and molecular emission
is the primary method for obtaining information on kinematic
motions inside molecular clouds.  Extensive efforts have been placed towards
using molecular tracers, typically CS and \HCOp , to search for the presence of
star-forming infall due
to gravitational collapse (see Evans 1999; Myers, Evans, \& Ohashi 2000 and
references therein).  The freeze-out of molecules onto
grain surfaces significantly reduces the effectiveness of using molecular
emission as a tracer of motions.  Indeed the inside-out collapse models of Shu (1977)
predict that the highest infall speeds are found in the exact regions where
molecules are expected to freeze-out.

Despite these arguments it is only recently that detections of gas phase freeze-out 
have become commonplace.   
Until recently inherent difficulties in extracting molecular abundances
from gas phase emission prevented definitive detections of gas-phase freeze-out.
Previous studies assumed local thermodynamic
equilibrium (LTE) to estimate total column densities. 
However, molecular emission from
tracers with high dipole moments is far from LTE.
In addition, 
secondary tracers such as CO and its isotopic variants were previously
used to estimate the total H$_2$ column density, since it cannot
be observed directly.  It is now known that depletion of CO in cold
regions seriously 
hampers its utility as an estimator of the total hydrogen
column density.   

The recent advent of sensitive continuum and heterodyne arrays 
probing millimeter/sub-millimeter wavelengths,
along with wide-field infrared imaging devices, has led to an 
explosion of clear detections of
gas phase depletion/freeze-out
(Bacmann et al. 2002; Bergin et al. 2002, Tafalla et al. 2002, Hotzel et al. 2002,
Caselli et al. 2002, Jessop \& Ward-Thompson 2001, Kramer et al. 1999, Alves et al. 1999).
At this conference alone there are 8 contributions discussing gas-phase
depletions (Carey et al., Di Francesco et al., Feldman et al., Savva et al.,
Kontinen et al., Lai et al., Lee et al., and Peng et al., this volume).
Observations of dust in emission or absorption 
provide direct knowledge of the dust distribution.
In addition, since the dust column density and mass is correlated 
with the H$_2$ column density and mass (Hildebrand 1983; Gordon 1995),  
these observations provide the clearest information
to date on the spatial distribution of H$_2$ molecules.   Furthermore, with
some geometrical assumptions, the radial profiles of core density
can be constructed; these methods have greatly increased our knowledge of the
physical conditions throughout the star formation process (Andr\'e et al. 2000,
Alves, Lada, \& Lada 2001).
Knowledge of the density and column density distribution of H$_2$
also significantly aids the molecular observations in two ways.
(1) By indirectly confirming the location of the H$_2$ density and column density peak
and, (2) by providing the density profile which helps to unravel the
similar effects of density and abundance on excitation.
These advances have moved chemical analyses beyond surveys of objects that
show evidence for complicated chemistry (e.g. TMC-1, L134N) towards studies
of more centrally concentrated objects that are closer to collapse and 
star formation.

In this concise review we outline how combined studies of
dust emission/absorption with molecular
emission have improved our ability to estimate molecular abundances and,
in consequence, how our picture of dark cloud chemistry has been 
changed to one dominated by the effects of freeze-out.   
In the following (\S\S 2 \& 3)
we briefly outline our current theoretical understanding of gas-grain interactions.
Furthermore we discuss key observational efforts that demonstrate the systematic nature
of gas-phase freeze-out in condensed cores and how these observations can be
qualitatively understood by a coupling of gas-grain chemical models to sophisticated
radiative transfer models.
In \S 4 we discuss how our improved understanding opens the door towards
using astrochemistry as a new and potent tool to study the process of star formation.   
 
\section{Gas-Phase Freeze-Out: Theory}

The rate of deposition of a molecule in the gas phase onto a grain surface is

\begin{equation}
k_{freeze-out} = \pi a^{2}\overline{v}Sn_{gr}\;\;(s^{-1}).
\end{equation}

\noindent where $a$ is the grain radius, $\overline{v}$ is the
mean thermal velocity, $S$ is the sticking coefficient, and $n_{gr}$ the space
density of grains.  If we use ``classical'' grains with $a =$
1000 \AA\ and $n_{gr}$ = 1.3 $\times$ 10$^{-12}$n($H_{2}$)
cm$^{-3}$ (Spitzer 1978), then the timescale for a molecule to freeze-out
onto a grain surface is:

\begin{equation}
\tau_{freeze-out} \sim \frac{2 \times 10^9\;{\rm yr}}{Smn({\rm{H_2})}}
\end{equation}

\noindent where $m$ is the molecular weight and $S$ is the sticking coefficient.
CO is a typical tracer of molecular cores which have typical densities of n(H$_2$) $\sim
10^5$ cm$^{-3}$ (Evans 1999).  For a sticking coefficient of unity
the freeze-out timescale is $< 10^4$ year, well above the estimated time
for thermal evaporation ($\gg 10^{8}$ yr), but below estimated ages of 
$\sim~1-10$ Myr (Palla and Stahler 2000; Hartmann 2001; Lee and Myers 1999).  
Thus it is surprising that a gas-phase is observed
at all, instead the near total freeze-out of the gas phase would be expected
(e.g. Iglesias 1977).  This points to the existence of some non-thermal mechanism to
desorb or remove molecules from the grains and allow for the observed active
gas phase chemistry.

Various mechanisms have been identified, including spot heating due to 
cosmic-ray or X-ray impacts, chemical desorption, and others.
The reader is referred to Williams (1993) for more detailed discussion.
The strength of the molecule bond to the grain surface is a key parameter 
that determines the efficiency of evaporation.  Because low temperatures preclude the
breaking of chemical bonds,
molecules are not expected to be chemically bound to grain surfaces.
Instead the approaching molecule has an induced dipole interaction with the
grain surface or mantle and is bound through weaker
Van der Waals-London interactions called physical adsorption
(Kittel 1996).  Van der Waals interactions are proportional to the product
of the polarizabilities of the molecule and the nearby surface
species ($E_b \propto \alpha_{mol}\alpha_{surface}$).  Because of this property
some species are more tightly bound to the grain surface than others.
For example CS is expected to be more tightly bound to the grain surface
than CO or N$_2$.  In addition, theoretical analysis has suggested that 
N$_2$ has a weaker bond to a water ice surface than CO (Sadlej et al. 1995).

Models which have achieved the best
success in reproducing observations are those
that incorporate these assumptions regarding bond strengths,  
resulting in a desorption process that preferentially removes more weakly bound species
(Bergin \& Langer 1997; Charnley 1997;
Nejad \& Wagenblast 1999; Aikawa et al. 2001; Li et al. 2002).  
As as sample in Figure~1 we provide 
chemical abundances as a function of time for a cloud with a centrally concentrated
density profile.  This result uses the model for B68 discussed by Bergin et al. (2002)
where the density profile is that for a Bonnor-Ebert sphere near-equilibrium that
is kept constant with time as the chemistry evolves.  This is a simplification
as more realistic models couple the physical and chemical evolution
(Bergin \& Langer 1997; Aikawa et al. 2001; Li et al. 2002).  Nonetheless
it is illustrative of the observed effects.  

\begin{small}
%\begin{verbatim}
\begin{figure}
%\centering
\vspace{3.0in}
\special{vscale=40.0 hscale=40.0 hoffset=-15.0 voffset=-20.0 angle=0
psfile=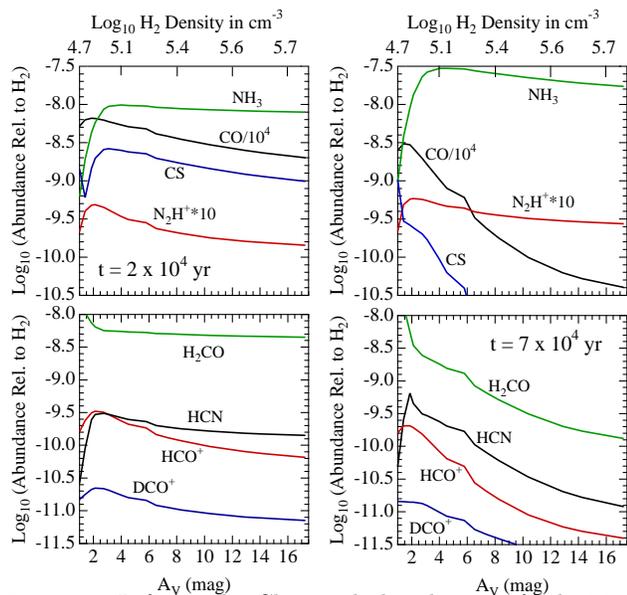}
\caption{
Left panels: Chemical abundances of select tracers as a function of cloud
depth (A$_V = 1$ to 17 mag) from a time dependent model sampled
at t = 2 $\times 10^{4}$ yr.   Right panels: same except for t = 7 $\times 10^{4}$ yr.
The top axis on left and right sides show the density that corresponds to
extinction on the lower x-axis.
The model shown is representative for the evolution of the chemistry in an
object that is centrally concentrated such that the highest densities are present
at the greatest extinctions.  This gives rise to the observed chemical structure as the
depletion rate increases with density. 
\label{fig-1}}
\end{figure}
%\end{verbatim}
\end{small}

Figure~1 shows chemical abundances for species that are typically used to
trace molecular clouds.  At early times (t = 2 $\times 10^{4}$ yr)
the effects of freeze-out are evident, but the overall
effect is small.  At later times (t = 7 $\times 10^{4}$ yr) strong
gas-phase depletions are seen for nearly all species.  CS has the largest
abundance decrease followed by other carbon-bearing species.  In contrast the nitrogen
chemistry remains active at later times due to the presumed volatility of
the N$_2$ molecule (see Bergin \& Langer 1997).  
One byproduct of CO depletion is an increase in the DCO$^+$/HCO$^+$ ratio
(see Aikawa et al. 2001; Caselli et al 2002; Caselli 2002 for further discussion).
In the following section 
we show that the pattern of differential
depletions -- early CS depletion, followed by CO, and finally N$_2$ -- is observed
in numerous sources.  Similar effects would be observed in any model that incorporates
these particular assumptions with regards to the gas-grain physics and evolves
to a centrally concentrated state. 

\section{Gas-Phase Freeze-Out: Observations}

Combined gas/dust investigations have resulted in the development of a new method 
to derive molecular abundances.   This method derives abundance profiles 
as a function of radius,
as opposed to a line of sight average.  More specifically, the observed core density
profile from the dust observations is used as an input to 
one-dimensional (1D) radiative transfer codes
(Monte-Carlo is generally used; e.g. Bernes 1979; Hogerheijde and van der Tak 2000).
The temperature profile is another
parameter  that is constrained by dust thermal emission or
by multi-transitional studies of molecular emission.  
%In practice the increase
%in density is large enough to dominate over small (less than a factor of two) changes
%in the temperature so, absent the presence of a star,
%constant temperature models are often appropriate.
Variables are the abundance profile and velocity structure (both bulk and
turbulent).  With these constraints, and iterating over the variables,
the radiative transfer model predicts the radial profile
of emission which can then be compared in detail to the observations.
%In general the shape of the abundance profile is constrained by the
%observed emission profile, but also through chemical considerations.
This technique has been outlined by van Dishoeck and Hogerheijde (1999)
and applications can be found in van der Tak et al. (2000), Tafalla et al. (2002),
Bergin et al. (2002), and Lee et al. (2002).

Below we discuss some recent investigations that highlight:
(1) the observed
depletions are differential in nature and are systemic in the dense ISM 
(Tafalla et al. 2002) 
and (2) current gas-grain
chemical models are capable of qualitatively matching observations of a limited
set of key species (Bergin et al. 2002). 

\subsection{Tafalla et al. 2002}

\begin{small}
%\begin{verbatim}
\begin{figure*}[t]
%\centering
\vspace{3.5in}
\special{vscale=65.0 hscale=65.0 hoffset=-30.0 voffset=330.0 angle=-90
psfile=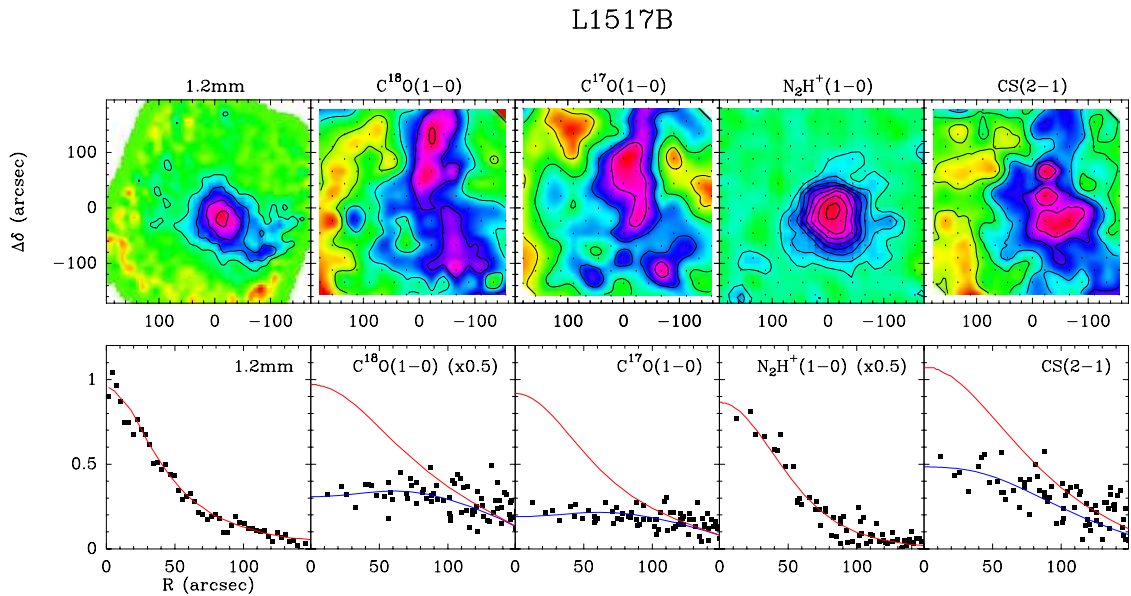}
\caption{
Top panels: Maps of 1.2mm continuum, {\rm C$^{18}$O}(1-0), {\rm C$^{17}$O}(1-0),
{\rm N$_2$H$^+$}(1-0), and {\rm CS}(2-1) emission in L1517B.
Bottom panels: Radial profiles of continuum and integrated emissions.
Observations are indicated by solid squares while the lines are results
from models of the radiative transfer for two abundance profiles.
The solid red lines are constant abundance models while the blue lines
represent models with a rapid decrease in the central abundance.
Taken from Tafalla et al. (2002).
\label{fig-double}}
\end{figure*}
%\end{verbatim}
\end{small}

In the top panels of Figure 2
we show integrated molecular emission maps of the centrally concentrated
molecular core L1517B
for sample species with a map of the 1.2 mm continuum emission (Tafalla et al. 2002).
The notable feature in this figure is the lack of correlation between the
1.2 mm continuum -- a tracer of H$_2$ -- with C$^{18}$O and CS emission; where
the continuum peaks the molecular emission exhibit local minima.   In contrast,
the N$_2$H$^+$ emission peak is coincident with the continuum emission maximum.
An analysis of these data using the physical description of the source
given by the dust continuum emission
is shown in the lower panels  (Tafalla et al. 2002).
Here the radial profiles of emission (solid squares) are compared to model predictions.
Models that assume constant abundance are shown as solid red lines; these models
predict excess emission at the core center for the CO isotopes and CS.
To reach agreement with observations  the abundances of these
species must be sharply reduced in the densest regions of the core (models shown
as solid blue lines) -- presumably through freeze-out.
Constant abundance models provide a reasonable match for N$_2$H$^+$,
while \NHthree\ requires a slight abundance increase towards
the center.  

Tafalla et al. also compare the observations directly to the chemical models
of Bergin \& Langer (1997), finding
good agreement to models of fast-collapse.  
Most significantly similar results are found for all 5 cores examined in this work,
as well as in sources observed by other studies
(Kuiper, Langer, \& Velusamy 1996; Ohashi et al. 1999; 
Lee et al. 2002; Bergin et al. 2002).
This demonstrates the systematic nature
of depletions -- they are observed in nearly all centrally concentrated
low mass cores.   {\em Thus differential depletions in cold dark clouds 
must be a chemical signature of the earliest stages of star formation. }

\begin{small}
%\begin{verbatim}
\begin{figure*}[t]
%\centering
\vspace{3.0in}
\special{vscale=65.0 hscale=65.0 hoffset=0.0 voffset=300.0 angle=-90
psfile=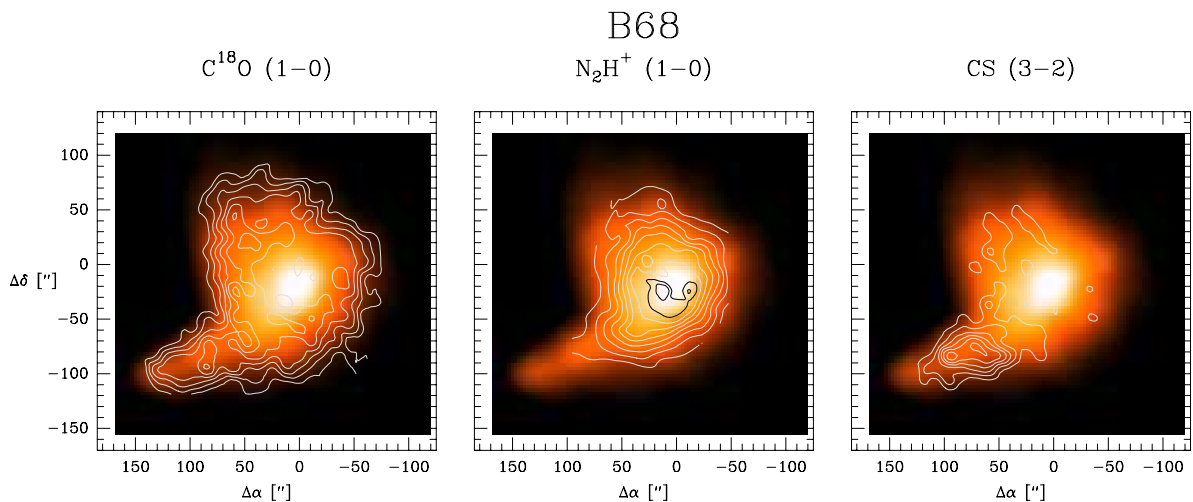}
\caption{
The observed emission distribution of visual extinction
or total column density (color/gray scale) in the B68 dark cloud.  Contours are
integrated emission maps of {\rm C$^{18}$O} (left-hand panel), {\rm N$_2$H$^+$}
(middle), and {\rm CS} (right-hand panel).  These observed emission distributions
show a sequence of increasing molecular depletion from {\rm CS}, to {\rm C$^{18}$O},
and even {\rm N$_2$H$^+$} (Bergin et al. 2002; Lada et al. 2002). 
\label{fig-double}}
\end{figure*}
%\end{verbatim}
\end{small}

\subsection{Bergin et al. 2002: B68}

The dark cloud B68 is another example of a centrally concentrated
object that exhibits emission features which are attributed to gas phase freeze-out.
In this core the H$_2$ distribution is traced through a map of dust visual extinction
(A$_V$
)
derived via near-infrared extinction mapping techniques (e.g.
Alves, Lada, and Lada 2001).   In Figure~3 we present the B68
A$_V$ map along with a series of
molecular emission maps.  In this core
we see that the N$_2$H$^+$ emission peaks in a shell partially
surrounding the peak of dust extinction.
Moreover, the N$_2$H$^+$ peaks inside the much larger C$^{18}$O emission hole, which
itself lies inside the CS emission depression.
Analysis of these data differed from Tafalla et al. (2002) in that a gas-grain chemical
model is directly linked to a radiative transfer model predicting the
profile of integration emission. This allows for chemical considerations to
determine how the abundance is structured with depth.  For example,
photodissociation limits the abundance at cloud edges and the competition
between depletion and desorption  determines how the abundance declines with
density.

Figure 4 presents an analysis of the N$_2$H$^+$ and C$^{18}$O emission; in this
case the open squares show 
the integrated emission as function of visual extinction.  
This particular representation of the data is independent of 
cloud geometry, depending only on physical variables.  Furthermore,
if the emission is in LTE and optically thin the integrated emission
from a source with constant abundance would appear as a straight line in this diagram
(e.g. Frerking et al. 1982; Lada et al. 1994).
Clearly the C$^{18}$O-A$_V$ relation deviates from a straight line at
high extinctions.  Such structure in this relation could be due to 
freeze-out or high opacities.  To discriminate between these
possibilities,  observations of a lesser abundant
isotope are typically needed (and in this case confirm freeze-out). 
The dashed line is the ``best fit'' radial abundance profile predicted in 
the chemical model and the solid line is the resulting emission profile.
For \CeiO\ the observations require significant freeze-out and also an
abundance reduction at low extinction due to photodissociation.  In the
case of \NtwoHp\ an abundance reduction at low extinction is 
required, but only a small decrease in abundance at high A$_V$ is needed to match
observations.

\begin{small}
%\begin{verbatim}
\begin{figure*}
%\centering
\vspace{2.5in}
\special{vscale=85.0 hscale=85.0 hoffset=0.0 voffset=-60.0 angle=0
psfile=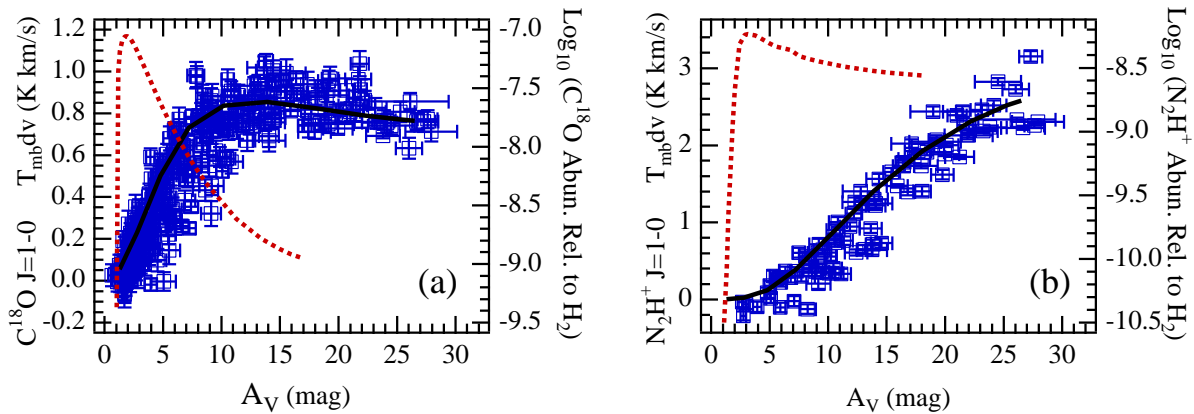}
\caption{
Point by point comparison of {\rm C$^{18}$O} J=1-0 and {\rm N$_2$H$^+$} integrated
intensity
with visual extinction for the entire B68 dark cloud.
In all plots the data are presented as open
squares with error bars while solid curves represent the emission predicted by
a model combining chemistry with a Monte-Carlo radiative transfer code.
The dashed lines are the best fit molecular abundance profiles with the
axis labeled to the right (abundance of {\rm N$_2$H$^+$} is multiplied
by 100).  In a symmetric model
the abundance profiles represent the radial chemical structure 
within a pencil beam towards the extinction peak 
and thus are shown only for one-half of the total maximum extinction. 
Adapted from Bergin et al. (2002).
\label{fig-double}}
\end{figure*}
%\end{verbatim}
\end{small}

\begin{small}
%\begin{verbatim}
\begin{figure*}
%\centering
\vspace{3.5in}
\special{vscale=65.0 hscale=65.0 hoffset=0.0 voffset=-60.0 angle=0
psfile=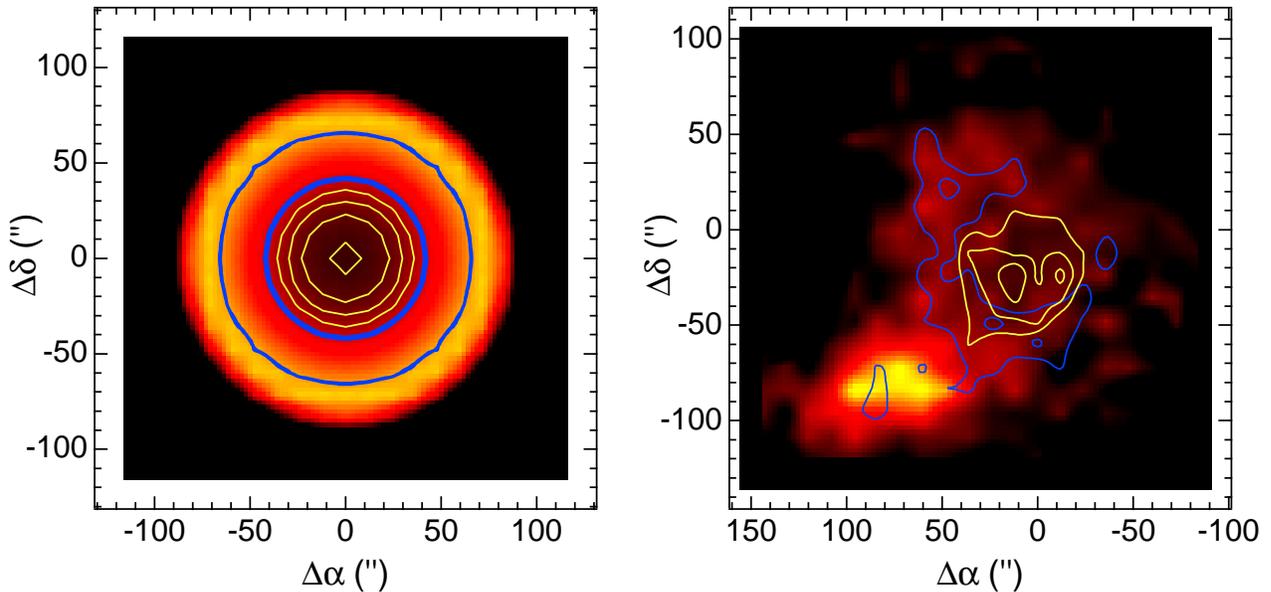}
\caption{
Comparison of model emission predictions with observational data.
Left: Color scale map of {\rm CS} (J=3-2) emission predicted from
the B68 chemical model.  Also shown are the peak contours of {\rm
C$^{18}$O} J=1-0 (blue) and {\rm N$_2$H$^+$} J=1-0  (yellow).  
Right: Similar plot of the observed emission in B68. 
Although the core clearly shows deviations from spherical symmetry there
is a close correspondence between model predictions and observations.
\label{fig-double}}
\end{figure*}
%\end{verbatim}
\end{small}

In Figure~5 we present a  more qualitative comparison of the structure
predicted by the chemical models to the B68 data.  
Here the left-hand panel represents the peak 
contours predicted by a spherical radiative transfer model combined
with the chemistry for transitions of CS, \CeiO\ and \NtwoHp .
The right-hand panel shows an identical representation of the observed
emission in B68.  Although clear divergences from spherical symmetry are
evident there exists
striking agreement between model and observation.
The results presented here and elsewhere suggests that we are 
making progress in our understanding of the chemistry
in these cold regions (e.g. Tafalla et al. 2002; Aikawa et al. 2001; Li et al. 2002;
Lee et al. 2002).  The decrease in the \NtwoHp\ abundance is also
significant as this suggests that its parent molecule  (N$_2$) is freezing onto grain
surfaces.  If this result is observed beyond B68 it could have important
implications for our ability to use molecules as the tracers of star formation
dynamics.

\subsection{Summary of Trends}

The above results are excellent examples of the trends that are becoming
apparent in chemical studies of low mass dark clouds.  These trends have
suggested changes in the way molecular emission is used to probe star forming
cores.

\begin{enumerate}
\item The use of C$^{18}$O as a probe of the H$_2$ column density in regions 
below the sublimation temperature of CO ($\sim 20$ K) is no longer appropriate.
Whether this extends to warmer cores in giant molecular
clouds remains an open issue (see Mauersberger et al. 1992; Chandler \& Carlstrom 1996; 
Gibb \& Little 1998; Savva et al., this volume)

\item 
Due to freeze-out the CS abundance can be sharply reduced when $n > 10^4$ cm$^{-3}$
and is no longer a reliable tracer of motions in high density gas.  However,
this molecule is still a good probe of low density molecular gas prior to condensation and
after star formation.

\item The emission from  N$_2$H$^{+}$ and NH$_3$
most closely resembles the maps of dust emission/extinction.
N$_2$D$^+$, DCO$^+$, and NH$_2$D are also found to be well correlated
with the dust (Caselli et al. 2002; Shah and Wooten 2001).
These tracers are emerging as the best probes
of the innermost regions of dense cores prior to the formation of a star.

\item Once a star has been born the heating of the inner envelope
returns some tracers to the gas phase.  This has not specifically been
addressed in this paper but the reader is referred to recent work in this
area including: Hogerheijde et al. (1999), Maret et al. (2002),
J{\o}rgensen et al. (2002), and references therein.
\end{enumerate}

\section{The Future of Dark Cloud Astrochemistry}

For a few key species we are now finding a convergence between
observation and theory.  This opens the door towards using chemistry
as a new tool to study the process of  star formation.   Below we discuss  
how this can be accomplished through both observational and theoretical
efforts.

\subsection{Observations}

One new tool is the isolation of specific tracers that either do not
freeze-out or freeze onto grain surfaces at late stages.  Observations
of these tracers (e.g. \NtwoHp ) over small or large scales
can provide complementary kinematical information to continuum observations.
For instance, in regions that are currently forming stellar clusters,
we can investigated the core-core velocity dispersion and dense core gas motions
relative to the overall structure (see Walsh; Hogerheijde, this volume).
On similar scales combined studies of an ``early'' depleter (e.g. CS)
and a high density tracer (e.g. \NtwoHp ) can also take advantage of the chemical
evolution to detect cores prior to condensation (CS),
during condensed stages (\NtwoHp ) and after star formation (see Lee et al. 2002;
Lee, this volume).  

The presence of differential depletions can certainly reduce the effectiveness
of using molecular emission as a tracer of star formation kinematics,
However, with careful choice of tracers one can use our expanded chemical
knowledge to reconstruct the radial profile of kinematical motions.   Indeed it 
is likely that {\em the key to fully unraveling star formation dynamics lies in our
improved understanding of the chemistry}.   One example is in B68 where 
the emission lines of \CeiO\ and \NtwoHp\ are characterized by a systematic
and well defined velocity gradient across the face of the cloud.  However, the
magnitude and direction of the velocity gradient is different between the two
tracers.  The chemical analysis presented in Figure~4 shows that \NtwoHp\ 
emission traces the innermost regions of the B68 core, while emission from
\CeiO\ predominantly traces the outer layers.   Thus, if the gradient is due
to rotation, then the combined chemical/dynamical analysis suggests that the
core is differentially rotating (Lada et al. 2002).

Another example of the importance of chemistry to dynamics is shown in
Figure~8.  
The left-hand panel  shows 
spectra of 3 molecules taken towards the same position in the L183 molecular core. 
The double peaked structure observed in the optically thick CS and HCO$^+$ emission
is now commonly used as a tracer of motions along the line of sight (e.g. Evans 1999).
However, in L183 the CS profile is consistent with infalling motions, while
HCO$^+$ is consistent with outflow.  
Since both of these transitions sample
the same level of excitation this set of conflicting data
must be due to chemistry. In particular, the answer must lie
in the differences in the abundance structure of each molecule, and changes
in the velocity field, as a function of depth.

\begin{small}
%\begin{verbatim}
\begin{figure*}
%\centering
\vspace{3.0in}
\special{vscale=55.0 hscale=55.0 hoffset=0.0 voffset=-40.0 angle=0
psfile=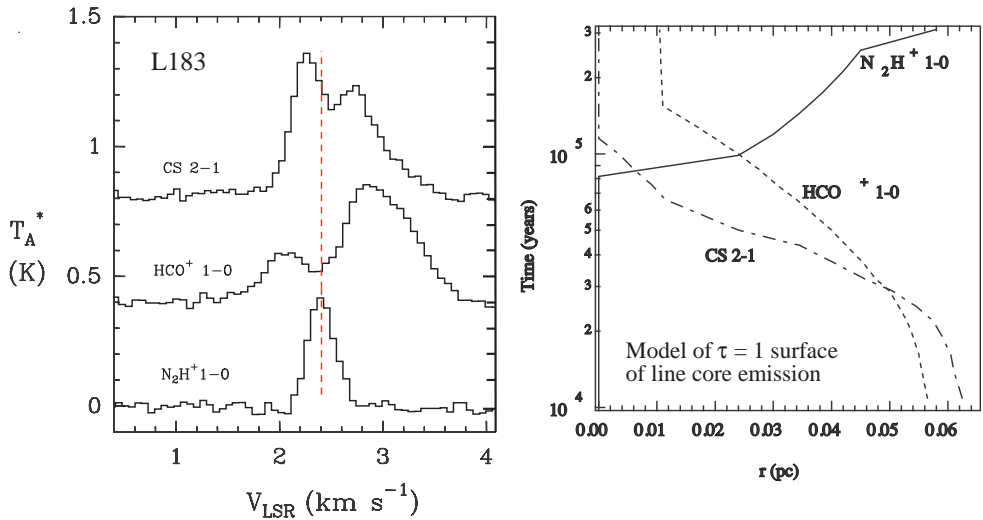}
\caption{
Left: Comparison of {\rm CS (2-1)}, {\rm HCO$^{+}$ (1-0)}, and {\rm N$_2$H$^+$ (1-0)}
spectra taken towards the same position in the  dark cloud L183.  The vertical
line shows the systemic velocity of the cloud as determined by the
optically thin {\rm N$_2$H$^+$} emission.  The {\rm CS} and {\rm HCO$^{+}$}
optically thick spectra show evidence for motions along the line
of sight.  However, for {\rm CS} the spectra is consistent with infall, while
{\rm HCO$^{+}$} appears to be consistent with outflowing motions (Lee \& Myers
2002, in preparation).
Right:
Plot showing the emission surface for selected molecular transitions
from a combined chemical and radiative transfer model of B68. 
The x-axis is the radius position in the core 
and the y-axis shows the chemical evolution.   
The lines delineate the $\tau = 1$ surface for the core of the spectral
line of selected molecular
transitions.  All positions to the left of a given line trace greater
opacities and are not probed by observations. 
When optically thick molecular 
line is used to trace motions via asymmetries in the profile the emission
preferentially probes the $\tau = 1$ surface and hence the motions at that
depth.   This plot demonstrates that as the chemistry evolves each 
molecule is sensitive to motions at different regions along the line of 
sight.  
\label{fig-double}}
\end{figure*}
%\end{verbatim}
\end{small}
A qualitative example of how these tracers could sample different 
motions along the line of sight is shown in the right-hand panel of Figure~8.
This example uses the  
combined chemical/radiative transfer model constructed for B68.  In the plot
the x-axis is the radial position inside the cloud with a radius of
0.065 pc, and the y-axis shows the time evolution.  The lines delineate
the $\tau = 1$ surface of the line core emission.  All positions to the left of a given
line for a particular transition have higher opacities.   
For example, for t~$< 8 \times 10^4$~yr emission from  \NtwoHp\  is optically
thin and traces the entire line of sight through the cloud.
At later times the abundance
rises and the emission becomes thick and only probes a surface that is
successively further from the core center.  

When optically thick molecular transitions are used to probe motion via 
asymmetric self-reversals the emission
preferentially probes the $\tau = 1$ surface and hence the magnitude
and direction of motions at that depth. 
{\em This plot clearly demonstrates that as the chemistry evolves the various tracers are
sensitive to different layers within the cloud.}  For instance
in this model
CS (J=2-1) and HCO$^+$ (J=1-0) initially both probe the outer layers.  
As the cloud evolves, CS differentially depletes
and the $\tau = 1$ surface retreats deeper into the cloud.  
Eventually, (for t~$> 3 \times 10^4$~yr)
HCO$^{+}$ emission traces outer layers, while CS probes the inner layers. 
This plot is not supplied as a template for the actual
evolution of any particular core, rather it is shown 
as an example of how evolutionary changes in the chemical
composition can affect observed emission, and more importantly the
interpretation. Therefore, with observation of several species, 
and knowledge of the chemical abundance and physical profiles,
another powerful new tool is the ability to search for structure 
in the radial velocity field, a key to understanding
star formation.

\section{Theory}

Our understanding of chemical theory in dark clouds also must evolve.
Areas that clearly need further investigation
include comparison of models to observations of similar cores
(L1544, L1517B, B68) in species beyond the typical
suite of tracers (e.g. CO, CS, \NtwoHp ).  For example Lee et al. (2002) find
that the abundance of CCS is depleted in the the L1512 and L1544 cores, but
find evidence for a small abundance rise in the center.  This would be in
agreement with the models of Ruffle et al. (1997) that suggest a rebound
in the abundances of complex molecules due to CO freeze-out. 
Do other complex molecules provide similar results and can we further
refine our understanding of gas phase pathways?    This moves
the testing of chemical models beyond comparison to abundances
derived in template sources, such as TMC-1 and L134N, to 
observational comparisons with more centrally condensed objects that have well
defined physical properties.    Abundances derived in these sources 
still have uncertainties due to calibration, collision rates, and geometrical
assumptions, but
can generally be considered more reliable than simple line of sight averages.  
Finally one challenge for future theory is to place the above
observations in context of the non-detections of  
water vapor and molecular oxygen in cold cores (Snell et al. 2000;
Bergin \& Snell 2002).   Some efforts have begun to examine this
question (Viti et al 2001; Charnley et al 2001; Roberts \& Herbst 2002).

Models that combine chemistry with dynamics have become increasingly
more sophisticated.  The ultimate goal of these efforts is to use 
chemical observations to refine our understanding of star formation
theory.   This goal is noteworthy and it remains to be seen whether the
chemical unknowns regarding gas-grain interactions dominate (e.g. sticking coefficients,
binding energies, grain sub-structure, surface chemistry, exact
desorption process).   
Combined models have recently begun to attempt to reproduce not
only the pattern of depletions, but also the column densities observed
in L1544, a new template source (Aikawa et al. 2001; Li et al. 2002).
This commendable step is needed to 
move theory beyond qualitative towards
quantitative comparisons with observations.  
However, in \S 3.2 we demonstrated that we now have 
the capability to place model predictions into radiative transfer
models and predict direct observables.  This is important because 
column density estimates depend on assumptions regarding excitation that may
not be reproduced in the model at the given time where model column densities
best match observational estimates.  Hence one could reproduce the column density
and not the observed emission because of differences
in the radial profile of physical conditions.  This task is time consuming,
but sophisticated radiative transfer codes are now readily available 
(see Hogerheijde \& van der Tak 2002).
The combination of chemical/dynamical models with radiative transfer 
truly offers the best opportunity to take chemical models and astrochemistry
to the next level and potentially begin to test star formation theory.

\section{Summary}

The chemical processes that result in the formation of molecules play an 
important role in star and planetary formation through molecular contributions to 
the stability of the cloud against gravitational forces.  Given the widespread
nature of gas-phase freeze-out it is becoming increasingly evident that
knowledge of the chemistry is also required in order to fully 
characterize the star formation process.   
The future of dark cloud chemical studies is promising 
as more complex and predictive models will be developed that can 
be compared to observations with increasing resolution and sensitivity.
This will lead to crucial tests of chemical theory but also
presents a completely new approach to star formation studies,
bringing astrochemistry into a wider realm as a potent tool. 

\acknowledgements

We are grateful to I. Bergin for proofing the manuscript
and M. Tafalla for providing comprehensive figures
for this presentation.  We are also grateful for our collaboration with 
Charlie Lada, Jo\~ao Alves, and Tracy Huard which made the B68 analysis possible.

\end{document}